\documentclass[a4paper,onecolumn,10pt,noarxiv,accepted=2025-05-23]{quantumarticle}\pdfoutput=1

\usepackage{graphics}
\usepackage{amssymb}
\usepackage{amsmath}
\usepackage{epsfig}
\usepackage{color}
\usepackage{tabu}
\usepackage{mathtools}
\usepackage[colorlinks,linkcolor=blue,anchorcolor=blue,citecolor=blue,urlcolor=blue]{hyperref}
\usepackage{physics}
\usepackage{float}
\usepackage{tikz}
\usetikzlibrary{quantikz}
\usetikzlibrary{calc,positioning}
\usepackage{bm}

\newcommand{\casql}{Laboratory of Quantum Information, University of Science and Technology of China, Hefei, Anhui, 230026, China}
\newcommand{\casex}{CAS Center For Excellence in Quantum Information and Quantum Physics, University of Science and Technology of China, Hefei, Anhui, 230026, China}
\newcommand{\hfnational}{Hefei National Laboratory, University of Science and Technology of China, Hefei, Anhui, 230088, China}
\newcommand{\aihf}{Institute of Artificial Intelligence, Hefei Comprehensive National Science Center, Hefei, Anhui, 230088, China}
\newcommand{\origin}{Origin Quantum Computing, Hefei, Anhui, 230088, China}

\begin{document}

\title{Simulation of open quantum systems on universal quantum computers}

\author{Huan-Yu Liu}
\orcid{0000-0002-6158-9627}
\affiliation{\casql}
\affiliation{\casex}

\author{Xiaoshui Lin}
\affiliation{\casql}

\author{Zhao-Yun Chen}
\orcid{0000-0002-5181-160X}
\affiliation{\aihf}

\author{Cheng Xue}
\orcid{0000-0003-2207-9998}
\affiliation{\aihf}

\author{Tai-Ping Sun}
\orcid{0009-0009-2591-1672}
\affiliation{\casql}
\affiliation{\casex}

\author{Qing-Song Li}
\orcid{0000-0002-8133-4939}
\affiliation{\casql}
\affiliation{\casex}

\author{Xi-Ning Zhuang}
\orcid{0000-0001-5118-5066}
\affiliation{\casql}
\affiliation{\casex}
\affiliation{\origin}

\author{Yun-Jie Wang}
\orcid{0009-0000-5347-5286}
\affiliation{Institute of the Advanced Technology, University of Science and Technology of China, Hefei, Anhui, 230088, China}

\author{Yu-Chun Wu}
\orcid{0000-0002-8997-3030}
\email{wuyuchun@ustc.edu.cn}
\affiliation{\casql}
\affiliation{\casex}
\affiliation{\aihf}

\author{Ming Gong}
\orcid{0000-0003-1055-7147}
\email{gongm@ustc.edu.cn}
\affiliation{\casql}
\affiliation{\casex}
\affiliation{\hfnational}

\author{Guo-Ping Guo}
\orcid{0000-0002-2179-9507}
\email{gpguo@ustc.edu.cn}
\affiliation{\casql}
\affiliation{\casex}
\affiliation{\aihf}
\affiliation{\origin}

\begin{abstract} 
The rapid development of quantum computers has enabled demonstrations of quantum advantages on various tasks. However, real quantum systems are always dissipative due to their inevitable interaction with the environment, and the resulting non-unitary dynamics make quantum simulation challenging with only unitary quantum gates. In this work, we present an innovative and scalable method to simulate open quantum systems using quantum computers. We define an adjoint density matrix as a counterpart of the true density matrix, which reduces to a mixed-unitary quantum channel and thus can be effectively sampled using quantum computers. This method has several benefits, including no need for auxiliary qubits and noteworthy scalability. Moreover, some long-time properties like steady states and the thermal equilibrium can also be investigated as the adjoint density matrix and the true dissipated one converge to the same state. Finally, we present deployments of this theory in the dissipative quantum $XY$ model for the evolution of correlation and entropy with short-time dynamics and the disordered Heisenberg model for many-body localization with long-time dynamics. This work promotes the study of real-world many-body dynamics with quantum computers, highlighting the potential to demonstrate practical quantum advantages.
\end{abstract}

\maketitle

\section{Introduction}

%Georgescu2014quantumsimulation,Minganti2018theorydpt,Cleve2016Efficient,Manzano2020lindbladme
% The concept of quantum computers inaugurated by Feynman in his famous talk to simulate quantum systems \cite{Hey2023Book, nielsen2010quantum}, which can not be efficiently achieved with classical computers due to the exponential growth of the Hilbert space with the increasing system size, has been the major catalyst for quantum computation in the past forty years.
The concept of quantum computers, introduced by Feynman in his famous talk on simulating quantum systems \cite{Hey2023Book, nielsen2010quantum}, has driven the development of quantum computation over the past forty years. This approach addresses the inefficiency of classical computers in simulating quantum systems, where the Hilbert space grows exponentially with system size. To date, quantum processors with high-performance superconducting qubits  \cite{Kim2023evidence}, photons \cite{Deng2023Gaussian}, or reconfigurable neutral atom arrays \cite{Lukin12023Logical}, have been realized for quantum advantages \cite{Daley2022Practical, Huang2022quantumadvantage,Trivedi2024Quantum,Diego2017Demonstration}. With efficient implementable unitary quantum gates, simulating quantum systems with these hardware has become a stirring research forefront \cite{Kim2023evidence, Trivedi2024Quantum, Fauseweh2024quantumsimulation, Bauer2023quantumsimulation, Fauseweh2024Quantum}. However, real quantum systems are inevitably dissipated from the interaction with the environments, leading to decoherence and relaxation processes \cite{ISAR1994opensystems, Preskill2018quantumcomputingin, Kenneth2024Usingml}, thus quantum error mitigation \cite{Cai2023quantumem, van2023probabilistic, Brien2023Purification, Youngseok2023Scalable} and quantum error correction \cite{Campbell2017roadsftqc, Babu2023errorcorrection, Sivak2023realtime, Miao2023overcoming} strategies should be employed for practical quantum computations. Alternatively, quantum noise and the associated dynamical process can lead to Intriguing and fundamental physical phenomena like dissipative phase transition \cite{Kessler2012dissipativeptspin, Fitzpatrick2017observationdpt, Lyu2024Multicritical} and topological protection \cite{Yao2021topological, deGroot2022symmetryprotected, Paszko2024Edge}. In these cases, the system follows non-unitary dynamics, and thus can not be naturally simulated on the quantum hardware. Therefore, efficient simulation of open quantum systems with quantum computers has become an indispensable open question.

There are already several strategies for simulating open quantum systems on quantum computers \cite{rmpWeimer2021Simulation, delgado2024quantum}. For instance, the most straightforward way is to approximate the non-unitary operator with some implementable unitary operations \cite{Kamakari2022Digital}. The second method is to encode the non-unitary operation into a unitary form in the extended Hilbert space \cite{Schlimgen2021Quantum, han2021experimental, Alexander2020bath,de2021quantumheavyion, Ding2024simulating, Wang2011quantum}, in which auxiliary qubits are required to construct the extended space. Some of these methods, like approximating the non-unitary dynamics with unitary ones and constructing the dilation, do not have well-established scalability. The third approach is to convert this problem into an optimization problem, which can then be solved using variational quantum algorithms \cite{Yuan2019TheoryVariational, Mahdian2020variational, Luo2024variational}. However, its practicability and capability for quantum advantages are still unclear \cite{Palma2023limitationvqa, Bittel2021trainingvqa, liu2023variational}. In experiments, the effectiveness of these strategies in real quantum chips has been demonstrated using a few qubits \cite{Schlimgen2021Quantum, han2021experimental, Kamakari2022Digital}. 

Here, we present a scalable approach to simulate open quantum system dynamics. Our idea comes from the mixed-unitary quantum channel \cite{Hann2021Resilience, Lee2020Detecting, Peetz2024simulation}, in which every Kraus operator is proportional to a unitary operation. This channel can be simulated by post-processing the results on quantum computers. Our central idea is to define an adjoint channel to approximate the dynamics of the dissipative system. The adjoint channel, defined to be mixed-unitary, can be efficiently simulated on quantum computers. This method has several advantages, including the elimination of auxiliary qubits, extraordinary scalability, and enabling long-time simulation. With this approach, we demonstrate two intriguing applications, one is the short-time dynamics of the correlation and entropy in the quantum $XY$ model, and the other is the long-time many-body localization (MBL) \cite{Alet2018manybodylocalization, Abanin2019Colloquiummbl} in the disordered Heisenberg model. This method has some extent of universality and significant application of this approach include quantum relaxation \cite{Zhang2020Anomalous, Fischer2016dynamicsmbl}, quantum error mitigation \cite{Cai2023quantumem, van2023probabilistic, Brien2023Purification, Youngseok2023Scalable}, and quantum error correction \cite{Campbell2017roadsftqc, Babu2023errorcorrection, Sivak2023realtime, Miao2023overcoming}, etc.

\section{Dynamics in open quantum systems}

In an open quantum system, the interaction between the system and its environment causes dissipation and relaxation. The corresponding dynamics can be described by the Lindblad master equation \cite{lindblad1976semigroups, GKSL1976,  Dariusz2017Historygksl, Manzano2020introductionlindblad}
\begin{equation}
\label{eq-lindblad}
    \frac{d\rho}{dt} = -i[H,\rho] + \sum_{k=1}^D \gamma_k \left(P_k\rho P_k^\dagger - \frac 12 \{ P_k^\dagger P_k,\rho\}\right),
\end{equation}
where $H$ and $\rho$ are the Hamiltonian and the density matrix of the system, $P_k$ and $\gamma_k$ are the dissipative operator and strength, $D$ is the number of dissipative terms, $[A, B]$ and $\{A, B\}$ refer to commuting and anti-commuting operations, respectively. Our central goal is to simulate this equation using a quantum computer. To this end, we define the dissipative operator to be a tensor product of Pauli operators as $P_k \in \{X, Y, Z, I \}^{\otimes n}$, with $n$ the number of qubits. This dissipative model has been ubiquitously used in research \cite{Zhang2020Anomalous, Fischer2016dynamicsmbl,
Cai2023quantumem, van2023probabilistic, Brien2023Purification, Youngseok2023Scalable,
Campbell2017roadsftqc, Babu2023errorcorrection, Sivak2023realtime, Miao2023overcoming, Guimar2023noiseassisted, Guimar2024optimized, Chen2018simulate}.

The evolution of the density matrix can be expressed as
\begin{equation}
\label{1orderdt}
    \rho(t+\delta t) = \rho(t) + \frac{d\rho(t)}{dt} \delta t + \mathcal{O}(\delta t^2),
\end{equation}
where $\delta t$ is the time-step. For convenience, hereafter, we set $t_0=0$ and define $\rho_x=\rho(x\delta t),x \in \mathbb{Z}$. The aforementioned dissipative process can be viewed as a quantum channel $\mathcal{E}$ with an operator-sum representation as  
\begin{equation}
\label{lindbladkraus}
    \rho_{x+1} = \mathcal{E}(\rho_x) = \sum_\alpha M_\alpha \rho_x M_\alpha^\dagger.
\end{equation}
The Kraus operators admit the following form \cite{preskill, Hu2022generalquantum, Zhang2023Quantum}
\begin{equation}
\label{eq-malpha}
    M_0 = I + (-iH+K) \delta t, \quad M_k = \sqrt{\delta t \gamma_k} P_k,
\end{equation}
where $k\in \{1,2\cdots,D\}$, $K=- \frac 12 \sum_k \gamma_k P_k^\dagger P_k$ such that $ \sum_\alpha M_\alpha^\dagger M_\alpha = I + \mathcal{O}(\delta t^2)$. 

\begin{figure}[ht]
    \centering
    \begin{tikzpicture}
    [lnode/.style={rectangle,draw=blue!30,fill=blue!10,align=center,very thick,minimum size=8mm},
    rnode/.style={rectangle,draw=green!30,fill=green!10,align=center,very thick,minimum size=8mm}]

    \node[rnode](rhot0) at (0,0){$\rho_0$  };
    \node[rnode](rhot1) at (2,0){$\rho_1$  };
    \node[rnode](rhot2) at (4.5,0){$\rho_2$  };
    \node[rnode](rhotk) at (7,0){$\cdots$  };

    \node[lnode](sigmat1) at (2,2){$\mathcal{F}(\rho_0)$};
    \node[lnode](sigmat2) at (4.5,2){$\mathcal{F}^2(\rho_0)$};
    \node[lnode](sigmatk) at (7,2){$\cdots$  };

    \draw (rhot0) -- (0,2);
    \draw [->] (0,2) -- node[above]{$\mathcal{F}$} (sigmat1);
    \draw [->] (sigmat1) -- node[above]{$\mathcal{F}$} (sigmat2);
    \draw [->] (sigmat2) -- node[above]{$\mathcal{F}$} (sigmatk);
    \draw [->] (rhot0) -- node[above]{$  \mathcal{R}  $}  (rhot1);
    \draw [line width=1.5pt,->] (rhot1) -- node[above]{$\mathcal{R} $}  (rhot2);
    \draw [line width=1.5pt,->] (rhot2) -- node[above]{$\mathcal{R} $}  (rhotk);
    \draw [->] (sigmat1) -- node[right]{$\mathcal{R} $} (rhot1);
    \draw [->] (sigmat2) -- node[right]{$\mathcal{R} $} (rhot2);
    \draw [->] (sigmatk) -- node[right]{$\mathcal{R} $} (rhotk);

    \node[scale=1] at (3.8,5) {
        \begin{tikzcd}
            \lstick{$\rho$} & \gate{V_1} & \meter{O} & \rstick[4]{
                $
                \begin{aligned}
                    \bm{s} = \{ \operatorname{Tr} [V_i \rho V_i^\dagger O] \}_{i=1}^M, \\
                    \operatorname{Tr}[\mathcal{E}_U(\rho)O ] \leftarrow \bar{\bm{s}}.
                \end{aligned}
                $
            } \\
            \lstick{$\rho$} & \gate{V_2} & \meter{O} & \\
            \lstick{$\vdots$} \\
            \lstick{$\rho$} & \gate{V_M} & \meter{O} &
        \end{tikzcd}
    };
    \node at (1.6,7.2) {$V_i \sim \{ p_i,U_i\}$};
    \node at (0,7) {\textbf{(a)}};
    \node at (0,2.5) {\textbf{(b)}};
\end{tikzpicture}
    \caption{
    {\bf Sampling of the mixed-unitary quantum channels and reconstruction of dissipated density matrix from the adjoint density matrix}. 
    (a) Method for sampling the mixed-unitary channel $\mathcal{E}_U(\rho)$ in Eq. \ref{eq-mixedchannel} on quantum computers. In each of the $M$ samples, apply $V_i$ on the initial state $\rho$ based on the probability distribution $\{p_i, U_i\}$ and then measure with the observable $O$. The target value is approximated by an average over all samples.
    (b) Workflow for the reconstruction of the dissipated density matrix $\rho_m$ using the adjoint density matrix. Blue and green boxes represent the quantum simulation part and classical reconstruction part, respectively. We apply the adjoint channel $\mathcal{F}$ successively on the initial state $\rho_0$ to obtain the adjoint density matrices $\{ \mathcal{F}(\rho_0), \mathcal{F}^2(\rho_0)  ,\cdots\}$. The dissipated density matrix $\rho_m$ can be reconstructed with Eq. \ref{rr} using $\mathcal{F}^m(\rho_0)$ and actual matrices $\{ \rho_0,\rho_1,\cdots, \rho_{m-1} \}$. Thick arrows mean that all the density matrices from its left are required in the reconstruction $\mathcal{R}$. 
    }
    \label{fig-workflowe}
\end{figure}
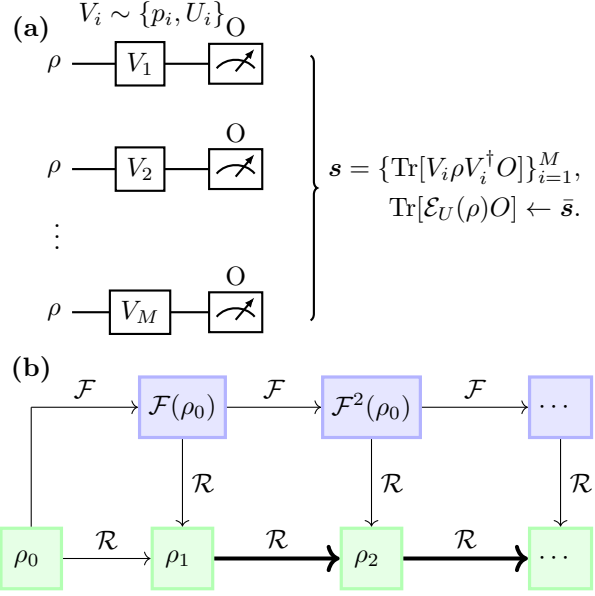

Generally, the above non-unitary channel can not be simulated directly on quantum computers. However, a special exception is the mixed-unitary quantum channel \cite{Hann2021Resilience, Lee2020Detecting, Peetz2024simulation} $\mathcal{E}_U$ with every Kraus operator satisfying $M_\alpha  \propto U_\alpha$, where $U_\alpha$ are unitary operations:
\begin{equation}
\label{eq-mixedchannel}
    \mathcal{E}_U(\rho) = \sum_\alpha p_\alpha U_\alpha \rho U_\alpha^\dagger, \qquad \sum_\alpha p_\alpha = 1.
\end{equation}
To simulate this channel, i.e, measure the evolved state $\mathcal{E}_U(\rho)$ with some observable $O$, we have
\begin{equation}
\label{eq-Eurho}
    \operatorname{Tr} [\mathcal{E}_U(\rho)O] = \sum_\alpha p_\alpha \operatorname{Tr}[U_\alpha\rho U_\alpha^\dagger O].
\end{equation}
This indicates that the simulation of this channel can be achieved via a sampling process shown in Fig. \ref{fig-workflowe}(a). In each sample, we apply $U_\alpha$ on the initial state with probability $p_\alpha$ and measure with operator $O$. The target value is averaged over all samples. With simple forms, mixed-unitary channels can exhibit many fundamental properties of general quantum channels \cite{Rosgen2008Additivity}, and have applications ranging from quantum cryptography \cite{Ambainis2000Private, Hayden2004Randomizing} to quantum encoding \cite{Hann2021Resilience}.

\section{The adjoint channel}\label{sec3}

Inspired by the capability of efficiently simulating  the mixed-unitary channel on quantum computers, we turn to transfer the general channel in Eq. \ref{eq-malpha} into an implementable form, called the adjoint channel
\begin{equation}
    \mathcal{F}(\rho)=\sum_\alpha N_\alpha \rho N_\alpha^\dagger.
\end{equation} 
Its output is called the adjoint density matrix, which is used to approximate and reconstruct the dissipated state. To this end, we define
\begin{equation}
\label{eq-sigma}
    \mathcal{F}^m(\rho_0) = \underbrace{\mathcal{F}\circ \mathcal{F} \circ \cdots  \circ\mathcal{F}}_{m} (\rho_0),
\end{equation}
which is the disjoint density matrix at $t = m\delta t$, and $\mathcal{F} \circ \mathcal{F} (\rho) = \mathcal{F} \left(\mathcal{F}(\rho)\right)$. 

The adjoint channel is set to be mixed-unitary as $N_\alpha \propto U_\alpha$, such that it can be sampled efficiently on quantum computers. Since $P_k$ is unitary, we already have $M_k \propto P_k,\forall k\in \{1,2,\cdots,D\}$. Consequently, when $k>0$, we set $N_k=M_k$.

Let $\sum_k \gamma_k=\Gamma$ and $P_k^\dagger P_k=I$, the term $K$ can be explicated as $K = -\frac 12 \Gamma I$. Then $M_0 = I - i H \delta t - \frac 12 \Gamma \delta t I$. To ensure $N_0\propto U_0$, we ignore its last term $ -\frac 12 \Gamma \delta tI$ and set $N_0=e^{-iH\delta t}=\mathcal{U}_{\delta t}$. Then $N_0$ equals the remaining terms up to an additive term of order $\mathcal{O}(\delta t^2)$:
\begin{equation}
    N_0 = \mathcal{U}_{\delta t} = I-iH\delta t + \mathcal{O}(\delta t^2), 
\end{equation}
To preserve the unit trace of density matrices, we need to re-normalize the channel as
\begin{equation}
\label{madjoint}
    N_0 = \frac{\mathcal{U}_{\delta t}}{ \sqrt{1+\Gamma \delta t}}, \quad N_k = \frac{\sqrt{\delta t\gamma_k }P_k}{ \sqrt{1+\Gamma \delta t}}.
\end{equation}
Then the adjoint density matrix yields 
\begin{equation}
\label{eq-adjointchannel}
    \mathcal{F}(\rho_0) =\frac{e^{-iH\delta t} \rho_0 e^{iH\delta t} + \sum_{k=1}^D \delta t\gamma_k P_k\rho_0 P_k}{1+\Gamma \delta t}. 
\end{equation}
This is one of the central expressions formulated in this work. Moreover, this kind of channel is also considered a ``parametric quantum channel" in studying quantum coherence \cite{matsoukasroubeas2023quantum}.

Now, the trace-preserving adjoint channel is in a mixed-unitary form, which can be simulated using quantum computers. However, compared to the quantum channel in Eq. \ref{lindbladkraus} that describes the dynamics of the quantum system, we ignored the term $-\frac 12 \Gamma\delta tI$ in $M_0$ and performed a renormalization procedure. Therefore, it is necessary to reconstruct the actual dynamics of the quantum system from the simulation results of the adjoint channel, which is shown in Sec. \ref{sec-reconstruct}.

\section{Reconstruction of the actual dynamical process}
\label{sec-reconstruct}

This section specifies the reconstruction of the dissipated density matrix from the simulated adjoint channel, including the time-evolved density matrix and its corresponding expectation values for the observable $O$. The workflow of the reconstruction process is shown in Fig. \ref{fig-workflowe}(b). In the following, we provide a brief explication of the results, and more details can be found in Appendix \ref{supp-functionr}.

Compare the difference between $\rho_1$ and $\mathcal{F}(\rho_0)$, while neglect terms of order $\mathcal{O}(\delta t^2)$, we have
\begin{equation}
\label{eq-refun1}
    \mathcal{F}(\rho_0) = \frac{1}{ 1+\Gamma \delta t  }\rho_1 + \frac{\Gamma \delta t}{1+\Gamma \delta t}\rho_0.
\end{equation}
Then $\rho_1$ can be reconstructed via $\rho_1 = (1 + \Gamma \delta t) \mathcal{F}(\rho_0) - \Gamma \delta t \rho_0$. The above result is correct for any $\rho_x$ as
\begin{equation}
\label{eq-refunm}
    \mathcal{F}(\rho_m) = \frac{1}{ 1+\Gamma \delta t  }\rho_{m+1} + \frac{\Gamma \delta t}{1+\Gamma \delta t}\rho_m.
\end{equation}
Since the quantum channels are linear transformations, we have 
\begin{equation}
    \mathcal{F}^{2}(\rho_0) = \frac{1}{ 1+\Gamma \delta t } \mathcal{F} (\rho_1)  + \frac{\Gamma \delta t}{1+\Gamma \delta t}  \mathcal{F} (\rho_0). 
\end{equation}
Apply this transformation successively, the adjoint density matrix becomes
\begin{equation}
    \mathcal{F}^m(\rho_0) = \frac{1}{ (1+\Gamma \delta t)^m  } \sum_{x=0}^{m} C_m^x (\Gamma \delta t)^{m-x} \rho_x,
    \label{eq-rhomreconstructed}
\end{equation}
where $C_m^x = \frac{m!}{x!(m-x)!}$. We can then directly reconstruct $\rho_m$ (see $\mathcal{R}$ in Fig. \ref{fig-workflowe} (b)) as
\begin{eqnarray}
\label{rr}
    \rho_m =(1+\Gamma \delta t)^m  \mathcal{F}^m(\rho_0) - \sum_{x=0}^{m-1} C_m^x (\Gamma \delta t)^{m-x} \rho_x.
\end{eqnarray}
This result means that reconstruction of the dissipated density matrix $\rho_m$ requires both the adjoint density matrix $\mathcal{F}^m(\rho_0)$, which is simulated on quantum computers, and the time-evolved states $\{ \rho_0,\rho_1,\cdots,\rho_{m-1} \}$, which have already been constructed in previous steps. This is another central formula obtained in this work. 
 
Furthermore, evaluating Eq. \ref{rr} is not accessible as the tomographic complexity for the density matrix will exponentially increase. Since the reconstructed density matrix is a linear combination of known states, using $ \operatorname{Tr} [ (\sum_i p_i\rho_i ) O ] =\sum_i p_i \operatorname{Tr} [ \rho_iO ]  $, the expectation value of the time-evolved expectation value can also be obtained with the same reconstruction procedure as:
\begin{equation}
    \operatorname{Tr} [\rho_mO] =  (1+\Gamma \delta t)^m  \operatorname{Tr} [\mathcal{F}^m(\rho_0)O]  -\sum_{x=0}^{m-1} C_m^x (\Gamma \delta t)^{m-x} \operatorname{Tr}[\rho_xO].
\end{equation}
Therefore, the proposed method can well characterize the dissipative dynamics and extract the physical observations.

A significant part of the reconstruction process is to simulate the effect of applying the adjoint channel $\mathcal{F}$ on the initial state $\rho_0$ multi-times, where the number of Kraus operators grows exponentially with evolution steps. Advanced sampling strategies like the quantum trajectory method \cite{Dalibard1992Wave, carmichael2009open} can accurately approximate the effect. Specifically, the error of $M$ samples scales $\mathcal{O}(1/\sqrt{M})$ according to the central limit theorem. This indicates that the sampling times do not need to increase with the number of steps. Therefore, the scalability of our method is guaranteed.

We now analyze the simulation error involved during evolution. Note that in Eq. \ref{eq-refun1}, we have neglected terms scaled as  $\mathcal{O}(\delta t^2)$, which is the error between the adjoint channel and the channel induced from the Lindblad master equation, which, according to Eq. \ref{1orderdt}, also has an approximation error scaled $\mathcal{O}(\delta t^2)$ as compared to its exact dynamics. Then the simulation error involved in one step with time-step $\delta t$ is $\mathcal{O}(\delta t^2)$. For multi-step evolution, in the worst case, errors will be accumulated in each step, resulting in a total simulation error scaled as $\mathcal{O}(m\times \delta t^2)=\mathcal{O}(T\delta t)$ for the overall simulation time $T=m\delta t$. Detailed error analysis can be found in Appendix \ref{supp-erroranalysis}.

For short-time simulations, this error scaling is acceptable. However, for long-time simulations, the error accumulation during the iteration may become significant. In this case, we can suppress $\delta t$ to reduce the simulation error. Furthermore, we prove that in the long time limit, both the adjoint density matrix and the dissipated density matrix will converge to the same steady state. To see this, assume that the steady state $\rho_s$ satisfying $d\rho_s/dt=0$ is reached at time $T$, thus we have $\rho_s = \rho(T)=\rho(T+\delta t)$. Then from Eq. \ref{eq-refun1} we have 
\begin{equation}
\label{eq-steady}
    \mathcal{F}(\rho_s)= \frac{\rho(T+\delta t)}{1+\Gamma \delta t}  + \frac{\Gamma\delta t\rho(T)}{1+\Gamma\delta t}  =\rho_s.
\end{equation}
Now it is clear that their steady states are the same. This provides the potential for long-time simulation in our procedure, especially when studying steady states \cite{Liu2021steady} and thermal equilibrium properties. Detailed data-processing analysis can be found in Appendix \ref{supp-dataprocess}. Moreover, this convergence property is also promising in some state preparation problems, like many-body thermal or ground states, which can be encoded as the steady states of Lindbladians \cite{Harrington2022Engineered, chen2023quantum, Ding2024Single, li2024dissipative, brunner2024lindblad}. With the adjoint channel, we can avoid simulating the exact dynamics, highlighting the potential of Lindblad engineering for efficient state preparation.

\section{Dissipative quantum $XY$ model}
\label{sec-xymodel}

\begin{figure}[ht]
    \centering
    \includegraphics[width=0.6\textwidth]{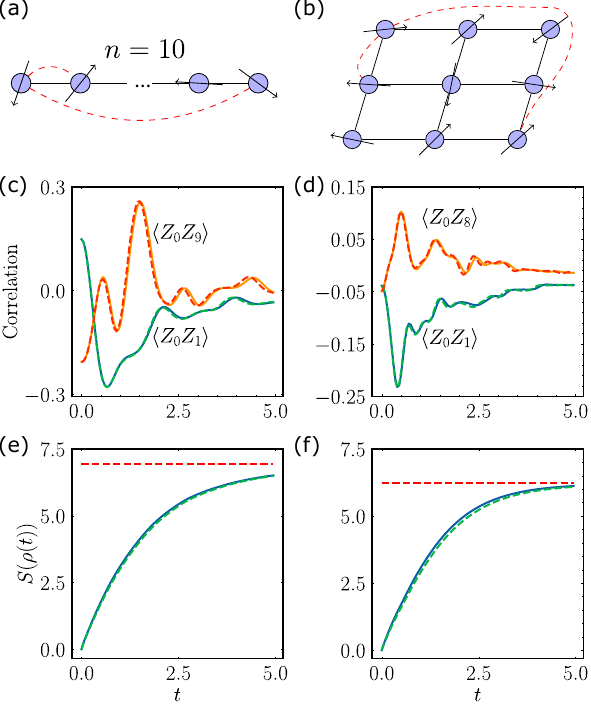}
    \caption{
    {\bf Simulation results for the dissipative quantum $XY$ model.}
    (a) - (b) The geometries of the 10-qubit one-dimensional $XY$ model and 9-qubit two-dimensional $XY$ model. The red dashed lines indicate the correlation to be calculated.  
    (c) - (d) The evolution of the corresponding correlation functions.
    (e) - (f) The evolution of the corresponding entropies. In (c) - (f), the dashed 
    lines are results from dissipated density matrices reconstructed from the adjoint density matrices; and the solid lines are results from the Lindblad equation without approximation. In (e) and (f), the horizon lines are entropies of the maximal mixed state $S(I/2^n)=n\ln 2$.
    }
    \label{fig-xymodel}
\end{figure}

Hereafter we provide numerical simulation results on several tasks to show both the credibility of this method and its potential in studying dissipated many-body models. We first consider the $n$-qubit quantum $XY$ model, whose Hamiltonian is
\begin{equation}
    H = -J \sum_{ \{i,j\} } (X_iX_j +Y_iY_j),
\end{equation}
where the $\{i,j\}$ represents nearest-neighbor interaction, with $J$ being its 
interaction strength, $X_i$ and $Y_i$ are the corresponding Pauli operator acting on the $i$-th qubit. For the $D=n$ dissipative terms, assume there is a Pauli-Z term on every qubit $P_k=Z_{k-1}$, $k\in \{1,2,\cdots, D\}$, with all dissipative strengths are set to be the same $\gamma_k=\gamma=0.1$. We study the 10-qubit one-dimensional model and 9-qubit two-dimensional model with size $3\times 3$, in which open boundary conditions are applied. The geometries are shown in Fig. \ref{fig-xymodel} (a) and (b), with $J = -1$ assumed in all simulations. 

We consider the short-time dynamics and the associated process to equilibrium, with total simulation time $T=5$ and time-step $\delta t = 0.05$. The initial state is a product state on all qubits $|\psi(0)\rangle =\bigotimes_{i=1}^n  (\cos \theta_i |0\rangle + \sin \theta_i |1\rangle )$ with each $\theta_i$ uniformly selected from $[0,2\pi]$. Eq. \ref{rr} is applied to reconstruct the dissipated density matrix. To characterize the dissipative process with $t\in[0,T]$, we study the evolution of the short-range correlation $\langle Z_0Z_1\rangle = \operatorname{Tr}[  Z_0Z_1 \rho(t)  ]$, long-range correlation $\langle Z_0Z_{n-1}\rangle$, and the von-Neumann entropy $S(\rho(t)) = -\operatorname{Tr} [  \rho(t) \ln \rho(t)  ]$. The package QuTiP \cite{qutip} is used to generate the actual dissipative process for comparison.
In contrast to the numerical calculations performed here, which aim to show the reliability of our method, information about quantum states on quantum chips can only be accessed by measurements. 
While the correlation function can be effectively measured, the acquisition of entropy is not the case since a costly tomography process could be needed.

The evolution of the correlation function is shown in Fig. \ref{fig-xymodel} (c) and (d); and the evolution of entropy are shown in Fig. \ref{fig-xymodel} (e) and (f). From these results, some important conclusions can be achieved. Firstly, the information obtained from our method shows strong correspondence with the exact process, which demonstrates the validity of our procedure. Secondly, we find that the short- and long-distance corrections do not show distinct distance-dependent features during evolution toward equilibrium. In the long-time limit ($t\in [5,10]$, more details can be found in Appendix \ref{supp-plott10}), we find that the correlation will approach zero due to the approaching of thermal equilibrium from ergodicity. Finally, from the assumption of ergodicity, the maximal mixed state $\rho = I/2^n$ will have entropy $S(\rho)=n\ln 2$, which is indeed approached in our simulation; see the two horizon solid lines in (e) and (f), independent of system dimensions. These results demonstrate that quantum computers can be used to study the fundamental physics in statistical physics in the many-body models for thermalization. 

\section{Many-body localization in dissipative Heisenberg model} 

Besides the above results, a more compelling avenue is to explore the MBL \cite{Alet2018manybodylocalization, Abanin2019Colloquiummbl}, which describes an isolated system without thermal equilibrium in the long-time evolution process. 
% This phenomenon has been widely studied and a direct connection between MBL and the random matrix theory has been achieved \cite{Buijsman2019randommatrix}. 
It has been studied with quantum computers both theoretically and experimentally \cite{Zhu2021probingmbl, Bauer2014analyzingmbl, Guo2021starkmbl}. To this end, we consider the one-dimensional Heisenberg model with the open-boundary condition and on-site disorders as
\begin{equation}\label{eq-xyzdisorder}
    H = -J \sum_{i=0}^{n-1} ( X_iX_{i+1} + Y_iY_{i+1} +Z_iZ_{i+1}  ) + \sum_{i=1}^n V_iZ_i,
\end{equation}
where $J$ is the nearest-interaction strength and $V_i\in [-h,h]$ is the disorder strength. Again, assume the dissipative term is a Pauli-Z on every qubit and $\gamma_k=\gamma$. In the simulation, we set $J=-1$ and $h=|10J|$. For long-time dynamics, $T=1000$ is considered, with time-step $\delta t = 0.05$. We simulate the 8-qubit model and the initial state is set as $|01010101\rangle$. To observe the effect of the dissipative strength, $\gamma$ is varied from $0$ to 1.

Following \cite{Fischer2016dynamicsmbl} and \cite{Alexander2019mbl}, we simulate the evolution of the von-Neumann entropy $S(\rho_L)$ and the imbalance $    I(t) = \frac{ \langle n_e-n_o\rangle  }{\langle n_e+n_o\rangle}$ in the dissipative process with $t\in [0, T]$, where $\rho_L$ is the reduced density matrix that is obtained by tracing out the half-sites $\rho_L = \operatorname{Tr}_{4,5,6,7} [\rho(t)]$, and $n_e$, $n_o$ are the occupation number on the even and odd sites, respectively. Note that the system is occupation number conserved, $[H,N_o]=0$, where the number operator can be expressed as $N_o=\sum_i n_i = \sum_i \frac{I-Z_i}{2}$ based on the Jordan-Wigner transformation \cite{Fradkin1989jwtransformation, li2022unified}. Then the imbalance can be re-written as $I(t) =  \frac 1n \sum_i  (-1)^i \langle Z_i\rangle$.

\begin{figure}[ht]
    \centering
    \includegraphics[width=0.6\linewidth]{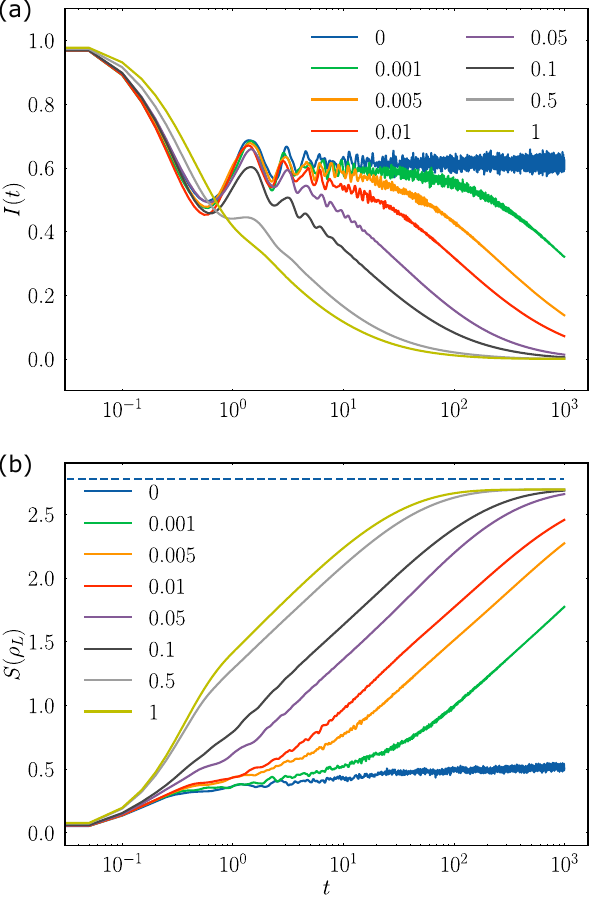}
    \caption{
    {\bf Simulation results for the dissipative Heisenberg model with on-site disorders}. 
    (a) and (b) are simulated evolution of imbalance $I(t)$ and entanglement entropy of $\rho_L$ in the dissipative process with $\gamma$ varied from 0 to 1. When $\gamma=0$, the isolated systems established the MBL without thermalization. Dissipation broke the  MBL, making the system reach thermal equilibrium. These results are in quantitative agreement with that in Refs. \cite{Fischer2016dynamicsmbl, Alexander2019mbl}.
    }
    \label{fig-imbalance}
\end{figure}

In the long-time simulation, we directly use the adjoint channel to approximate the actual dissipative process. The results are shown in Fig. \ref{fig-imbalance}. When $\gamma=0$, the entropy, and imbalance keep stable in the long-time simulation, showing that MBL is realized. In this regime, thermalization with ergodicity is not achieved. However, when dissipation is introduced, both the entropy and imbalance do not converge anymore due to thermalization. The larger $\gamma$ is, the quicker the thermalization will happen. In the dissipative models, the half-chain entropy will finally approach $S(\rho_L) = 4\ln 2 \approx 2.77$ from the maximal mixed state. All these results are in excellent agreement with previous analysis in \cite{Fischer2016dynamicsmbl} and \cite{Alexander2019mbl}, showing the flexibility of our method for simulating dissipative many-body models and their potential in the demonstration of quantum advantages.

\section{Discussion and outlook}

Quantum simulation of open quantum systems has become an important research forefront due to the development of quantum computers. Currently, many of the strategies for the simulation of open quantum systems do not possess well-established scalability, limiting their achievable simulation time and system size. Inspired by the simulatability of mixed-unitary quantum channels, we proposed a scalable method to simulate open quantum systems on universal quantum computers and demonstrated our method on two different dissipated many-body models. 

Several issues about applying this method to real quantum chips need to be considered in detail. Firstly, the most critical factor affecting the performance of current quantum computers is the decoherence noise, which greatly limits the depth of quantum circuits. Therefore, efficient error mitigation \cite{Cai2023quantumem, van2023probabilistic, Brien2023Purification, Youngseok2023Scalable} and circuit compilation \cite{Dmitri2017basiccc, Quetschlich2023predictcc, Kusyk2021Survey} strategies to mitigate the noise effect are required to improve the reliability of the results obtained on quantum computers. Secondly, mixed-unitary quantum channels are simulated via a sampling process. Note that the number of Kraus operators in the adjoint channel grows exponentially with the number of simulation steps. In Appendix \ref{supp-samplemixedchannel}, we present a brief illustration of flexibility in effectively sampling the adjoint channel. A more explicit sampling procedure, involving the sampling error analysis and circuit implementation scheme, is needed. Thirdly, we applied different data processing schemes in numerical simulations: using $\mathcal{R}$ in short-time simulations and directly using the adjoint channel in long-time simulations. Real simulations will include extra issues like sampling errors due to finite sampling measurement times or finite samples on the Kraus operators. A more flexible data-processing procedure to maintain the error may be desired. It is to be expected that the advancement of research related to open quantum systems could be realized with this method. We also hope that the capability of our method for long-time dynamics may have immediate application in some many-body models with slow relaxation 
\cite{Bordia2017Probing, Darkwah2022Probing, Dumitrescu2018Logarithmically}. 

\section*{Acknowledgement}

This work has been supported by the National Key Research and Development Program of China (Grant No.
2023YFB4502500, 2024YFB4504101), the National Natural Science Foundation of China (Grant No. 12404564, U23A2074), the Anhui Province Science and Technology Innovation (Grant No. 202423s06050001),  the Strategic Priority Research Program of the Chinese Academy of Sciences (Grant No. XDB0500000),  and the Innovation Program for Quantum Science and Technology (2021ZD0301200, 2021ZD0301500).

\section*{Data availability statement}

The data that support the findings of this study are available from the corresponding authors upon reasonable request.

% \clearpage
% \setcounter{table}{0}
% \renewcommand{\thetable}{S\arabic{table}}%
% \setcounter{figure}{0}
% \renewcommand{\thefigure}{S\arabic{figure}}%
% \setcounter{section}{0}
% \setcounter{equation}{0}
% \renewcommand{\theequation}{S\arabic{equation}}%
% % \setcounter{linenumber}{1}

% \onecolumngrid

\appendix
% \begin{center}
%     {\large \bf Supplementary Information for \\ ``Simulation of open quantum systems on universal quantum computers"}
% \end{center}

\section{Determine the reconstruction procedure}
\label{supp-functionr}

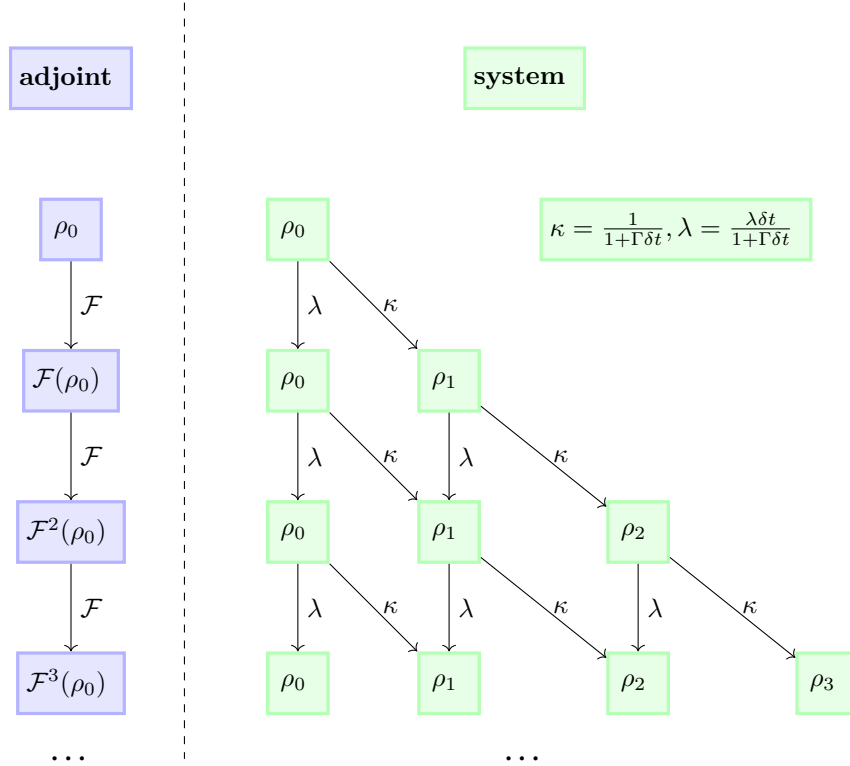
\begin{figure*}[ht]
    \centering
    \begin{tikzpicture}
    [lnode/.style={rectangle,draw=blue!30,fill=blue!10,align=center,very thick,minimum size=8mm},
    rnode/.style={rectangle,draw=green!30,fill=green!10,align=center,very thick,minimum size=8mm}]

    \node[lnode](rhol0) at (0,0){$\rho_0$  };
    \node[lnode](rhol1) at (0,-2){$\mathcal{F}(\rho_0)$  };
    \node[lnode](rhol2) at (0,-4){$\mathcal{F}^2(\rho_0)$  };
    \node[lnode](rhol3) at (0,-6){$\mathcal{F}^3(\rho_0)$  };

    \node[lnode](adjoint) at (0,2){ \bf adjoint  };

    \draw [->] (rhol0) -- node[right]{$\mathcal{F}$} (rhol1);
    \draw [->] (rhol1) -- node[right]{$\mathcal{F}$} (rhol2);
    \draw [->] (rhol2) -- node[right]{$\mathcal{F}$} (rhol3);

    \draw [dashed] (1.5,3) -- (1.5,-7);

    \node[rnode](rhom00) at (3,0){$\rho_0$  };
    \node[rnode](rhom01) at (3,-2){$\rho_0$  };
    \node[rnode](rhom02) at (3,-4){$\rho_0$  };
    \node[rnode](rhom03) at (3,-6){$\rho_0$  };

    \node[rnode](rhom11) at (5,-2){$\rho_1$  };
    \node[rnode](rhom12) at (5,-4){$\rho_1$  };
    \node[rnode](rhom13) at (5,-6){$\rho_1$  };

    \node[rnode](rhom22) at (7.5,-4){$\rho_2$  };
    \node[rnode](rhom23) at (7.5,-6){$\rho_2$  };

    \node[rnode](rhom33) at (10,-6){$\rho_3$  };

    \node[rnode] at (8,0){$\kappa = \frac{1}{1+\Gamma\delta t},\lambda=\frac{\lambda\delta t}{1+\Gamma\delta t}$  };

    \draw[->] (rhom00) -- node[right]{$\lambda$} (rhom01);
    \draw[->] (rhom01) -- node[right]{$\lambda$} (rhom02);
    \draw[->] (rhom02) -- node[right]{$\lambda$} (rhom03);
    \draw[->] (rhom11) -- node[right]{$\lambda$} (rhom12);
    \draw[->] (rhom12) -- node[right]{$\lambda$} (rhom13);
    \draw[->] (rhom22) -- node[right]{$\lambda$} (rhom23);

    \draw[->] (rhom00) -- node[right]{$\kappa$} (rhom11);
    \draw[->] (rhom01) -- node[right]{$\kappa$} (rhom12);
    \draw[->] (rhom02) -- node[right]{$\kappa$} (rhom13);
    \draw[->] (rhom11) -- node[right]{$\kappa$} (rhom22);
    \draw[->] (rhom12) -- node[right]{$\kappa$} (rhom23);
    \draw[->] (rhom22) -- node[right]{$\kappa$} (rhom33);

    \node[rnode](system) at (6,2){\bf system  };

    \node at (0,-7){$\bm{\cdots}$  };
    \node at (6,-7){$\bm{\cdots}$ };
\end{tikzpicture}
    \caption{
    {\bf Evolution of the component of the adjoint density matrix.}
    On the left side, we apply the adjoint channel $\mathcal{F}$ successively on the initial state $\rho_0$ to obtain the adjoint density matrix list. The right side shows the evolution of its components. Each arrow labeled with $\kappa/\lambda$ means the weight of the component. The diagram illustrates the relation of the trajectory between the truly evolved state and the adjoint one.
    }
    \label{figsupp_state}
\end{figure*}

Here we provide some more details on the reconstruction ($\mathcal{R}$ in Fig. \ref{fig-workflowe}(b)) of the dissipated density matrix from the simulated adjoint density matrix. Firstly, we expand the channel derived from the Lindblad master equation in Eq. \ref{eq-malpha}
\begin{equation}
    \begin{aligned}
       \rho_1 =& \sum_\alpha M_\alpha \rho_0 M_\alpha^\dagger \\
        =& \left( I - iH\delta t - \frac 12 \Gamma \delta t I  \right) \rho_0 \left( I + iH\delta t - \frac 12 \Gamma \delta t I  \right) + \sum_k \delta t \gamma_k P_k\rho_0 P_k \\
        =& (I-iH\delta t) \rho_0 (I+iH\delta t) - \Gamma \delta t \rho_0  +\sum_k \delta t \gamma_k P_k\rho_0 P_k + \mathcal{O} (\delta t^2).
    \end{aligned}
\end{equation}
Compare with Eq. \ref{eq-adjointchannel} and neglect terms scaled $\mathcal{O}(\delta t^2)$, we have
\begin{equation}
\label{eq-adac}
    \rho_1 = (1+\Gamma\delta t) \mathcal{F}(\rho_0) - \Gamma\delta t\rho_0,
\end{equation}
which can also be re-expressed as
\begin{equation}
\label{eq-r1function}
    \mathcal{F}(\rho_0) = \frac{1}{ 1+\Gamma \delta t  }\rho_1 + \frac{\Gamma \delta t}{1+\Gamma \delta t}\rho_0.
\end{equation}
Replace $t=0$ with $t=m\delta t$, the equation becomes
\begin{equation}
    \mathcal{F}(\rho_m) = \frac{1}{ 1+\Gamma \delta t  }\rho_{m+1} + \frac{\Gamma \delta t}{1+\Gamma \delta t}\rho_m.
\end{equation}
Then we can achieve long-time evolution based on the linear property of quantum channels. For instance, by applying the quantum channel $\mathcal{F}$ on the state $\rho_0$ twice we can obtain 
\begin{equation}
    \begin{aligned}
        \mathcal{F}^2(\rho_0) =& \mathcal{F} \left(\frac{1}{ 1+\Gamma \delta t  }\rho_1 + \frac{\Gamma \delta t}{1+\Gamma \delta t}\rho_0 \right) \\
        =& \frac{1}{1+\Gamma \delta t} \mathcal{F}(\rho_1) + \frac{\Gamma \delta t}{1+\Gamma \delta t} \mathcal{F}(\rho_0) \\
       =& \frac{1}{  (1+\Gamma\delta t)^2  } \rho_2 + \frac{2\Gamma \delta t}{ (1+\Gamma\delta t)^2 } \rho_1 + \frac{(\Gamma \delta t)^2}{ (1+\Gamma\delta t)^2 } \rho_0.
    \end{aligned}
\end{equation}
Based on this relation, when applying the channel successively, we have
\begin{equation}
\mathcal{F}^m(\rho_0) = \frac{1}{ (1+\Gamma \delta t)^m  } \sum_{x=0}^{m} C_m^x (\Gamma \delta t)^{m-x} \rho_x.
\end{equation}
The diagram for the evolution of the component of the adjoint density matrix is shown in Fig. \ref{figsupp_state}, which illustrates the relations of the trajectory between the adjoint density matrix and the actual ones.

With this relation, we are able to reconstruct  $\rho_m$ using collected information from $ \{ \mathcal{F}^m(\rho_0), \rho_{m-1},\cdots, \rho_0\}$ as 
\begin{equation}
    \rho_m =  (1+\Gamma \delta t)^m  \mathcal{F}^m(\rho_0) - \sum_{x=0}^{m-1} C_m^x (\Gamma \delta t)^{m-x} \rho_x.
\end{equation}
Besides the time-evolved density matrix, we usually need  to evaluate the expectation value of observable $O$ in real simulations. Note that in the above equation, $\rho_m$ is a linear combination of known states. Using $ \operatorname{Tr} [ (\sum_i p_i\rho_i ) O ] =\sum_i p_i \operatorname{Tr} [ \rho_iO ]$, we have
\begin{equation}
    \operatorname{Tr} [\rho_mO] =  (1+\Gamma \delta t)^m  \operatorname{Tr} [\mathcal{F}^m(\rho_0)O]  -\sum_{x=0}^{m-1} C_m^x (\Gamma \delta t)^{m-x} \operatorname{Tr}[\rho_xO].
\end{equation}
It can be seen that the time evolution of expectation values can be reconstructed with the same procedure. Therefore, simulation of open quantum systems can be achieved with this method. In the main text, we used this relation to reconstruct the dynamical process of the dissipative quantum $XY$ model.

\section{Error analysis}
\label{supp-erroranalysis}

Here we analyze the error between the dissipated density matrix and our reconstructed one. For simplicity, for the density matrix $\rho_m$, we denote the reconstructed density matrix as $\sigma_m$ in this part, provided with the initial state $\rho_0$.

We first bound the error in one step specified by the time-step $\delta t$. In Eq. \ref{eq-adac}. we neglect the error term scaled as $\mathcal{O}(\delta t^2)$, which is the error between the adjoint channel and the channel induced from the Lindblad master equation. The latter one, according to Eq. \ref{1orderdt}, also has an error scaled $\mathcal{O}(\delta t^2)$ compared to the exact dissipative process. Therefore, the error between the actual density matrix $\rho_1$ and our reconstructed one $\sigma_1 = (1+\Gamma \delta t) \mathcal{F}(\rho_0) -\Gamma \delta t \rho_0 $, denoted as $\epsilon_1$, is within $\mathcal{O}(\delta t^2)$:
\begin{equation}
    \rho_1 = \sigma_1 + \epsilon_1 = (1+\Gamma\delta t) \mathcal{F}(\rho_0) - \Gamma\delta t\rho_0 + \epsilon_1.
\end{equation}
Or equivalently,
\begin{equation}
    \mathcal{F}(\rho_0) = \frac{1}{1+\Gamma\delta t} \rho_1 + \frac{\Gamma\delta t}{1+\Gamma\delta t} \rho_0 - \frac{\epsilon_1}{1+\Gamma \delta t}.
\end{equation}
Assume $\mathcal{F}(\epsilon_x) \in \mathcal{O}(\delta t^2),\forall x$. When applying the adjoint channel $\mathcal{F}$ twice, we have
\begin{equation}
\begin{aligned}
     \mathcal{F}^2(\rho_0)  =& \frac{1}{1+\Gamma\delta t} \mathcal{F} (\rho_1) + \frac{\Gamma\delta t}{1+\Gamma\delta t} \mathcal{F} (\rho_0) - \frac{1}{1+\Gamma\delta t} \mathcal{F}(\epsilon_1) \\
     =& \frac{1}{1+\Gamma\delta t} \left(   \frac{1}{1+\Gamma\delta t} \rho_2 + \frac{\Gamma\delta t}{1+\Gamma\delta t} \rho_1 - \frac{\epsilon_2}{1+\Gamma \delta t} \right) \\
     &+ \frac{\Gamma\delta t}{1+\Gamma\delta t} \left(    \frac{1}{1+\Gamma\delta t} \rho_1 + \frac{\Gamma\delta t}{1+\Gamma\delta t} \rho_0 - \frac{\epsilon_1}{1+\Gamma \delta t}\right) 
     - \frac{1}{1+\Gamma\delta t}  \mathcal{F}(\epsilon_1)  \\
    =& \frac{1}{  (1+\Gamma\delta t)^2  } \rho_2 + \frac{2\Gamma \delta t}{ (1+\Gamma\delta t)^2 } \rho_1 + \frac{(\Gamma \delta t)^2}{ (1+\Gamma\delta t)^2 } \rho_0 - E,
\end{aligned}
\end{equation}
where
\begin{equation}
    E =  \frac{\epsilon_2}{(1+\Gamma\delta t)^2}  + \frac{\Gamma\delta t \epsilon_1}{(1+\Gamma\delta t)^2} + \frac{1}{1+\Gamma\delta t}  \mathcal{F}(\epsilon_1) .
\end{equation}
The  dissipated density matrix can be  reconstructed as
\begin{equation}
\label{eq-recon1}
    \sigma_2 = (1+\Gamma\delta t)^2 \mathcal{F}^2(\rho_0) 
    - 2\Gamma\delta t \rho_1 - (\Gamma\delta t)^2\rho_0 + (1+\Gamma\delta t)^2E.
\end{equation}
However, in real simulations, instead of $\rho_1$, we used the reconstructed state $\sigma_1=\rho_1-\epsilon_1$ in Eq. \ref{eq-recon1}. Therefore,
\begin{equation}
\begin{aligned}
    \sigma_2 =& (1+\Gamma\delta t)^2 \mathcal{F}^2(\rho_0) - 2\Gamma\delta t (\rho_1 -e_1) - (\Gamma\delta t)^2\rho_0 + (1+\Gamma\delta t)^2E \\
    =& (1+\Gamma\delta t)^2 \mathcal{F}^2(\rho_0) - 2\Gamma\delta t \rho_1- (\Gamma\delta t)^2\rho_0 + (1+\Gamma\delta t)^2E +2 \Gamma\delta t\epsilon_1
    \end{aligned}
\end{equation}
The error is
\begin{equation}
\begin{aligned}
   & (1+\Gamma\delta t)^2 \left(   \frac{\epsilon_2}{(1+\Gamma\delta t)^2}  + \frac{\Gamma\delta t \epsilon_1}{(1+\Gamma\delta t)^2} + \frac{1}{1+\Gamma\delta t}  \mathcal{F}(\epsilon_1) \right) + 2\Gamma\delta t \epsilon_1 \\
    =& \epsilon_2 + \mathcal{F}(\epsilon_1) + \Gamma\delta t[  \mathcal{F}(\epsilon_1)+2\epsilon_1  ].
\end{aligned}
\end{equation}
Since $\Gamma\delta t e_x\in \mathcal{O}(\delta t^3)$, which can be neglected. Then the above error $\epsilon_2 + \mathcal{F}(\epsilon_1)$ is within $\mathcal{O}(2\delta t^2)$.

Finally, following this, we can conclude that the simulation error at step $m$ is
\begin{equation}
     \mathcal{O} (m\delta t^2) = \mathcal{O} (T\delta t),
\end{equation}
where $T=m\delta t$ is the total simulation time. This error analysis is used in Sec. \ref{sec-reconstruct} in the main text. 

\section{Data processing strategies}
\label{supp-dataprocess}

In this section, we illustrate how to apply the reconstruction procedure while maintaining the accuracy. In the above, we have shown that the simulation error involved in the reconstruction procedure is $\mathcal{O}(\delta t^2)$. For short-time simulation, this reconstruction procedure works well, and the simulation results on the dissipative quantum $XY$ model, shown in Fig. \ref{fig-xymodel}, demonstrated the feasibility of this method. 

However, for long-time simulation, the reconstruction process can become unstable for two reasons. Firstly, the simulation error $\mathcal{O}(T\delta t)$ increases with the simulation time. Secondly, other factors, such as the sampling accuracy and circuit approximation in real simulations and their accumulation, may also become important. To see this, using the binomial expansion $(1+\Gamma\delta t)^m = \sum_{x=0}^{m} C_m^x (\Gamma\delta t)^{m-x}1^x$, we can re-write Eq. \ref{rr} as
\begin{align}
    \rho_m  =& (1+\Gamma \delta t)^m  \mathcal{F}^m(\rho_0)   - \sum_{x=0}^{m-1} C_m^x (\Gamma \delta t)^{m-x} \rho_x.\\
    =& \left(  \sum_{x=0}^{m} C_m^x (\Gamma\delta t)^{m-x}1^x \right) \mathcal{F}^m(\rho_0)   - \sum_{x=0}^{m-1} C_m^x (\Gamma \delta t)^{m-x} \rho_x.\\
    =& C_m^m \mathcal{F}^m(\rho_0) +   \sum_{x=0}^{m-1} C_m^x (\Gamma\delta t)^{m-x}  \mathcal{F}^m(\rho_0)   - \sum_{x=0}^{m-1} C_m^x (\Gamma \delta t)^{m-x} \rho_x.\\ 
    =& \mathcal{F}^m(\rho_0) + \sum_{x=0}^{m-1} C_m^x (\Gamma\delta t)^{m-x} \left[  \mathcal{F}^m(\rho_0) -\rho_x \right]. \label{eq-diff}
\end{align}
In the right-hand side of the Eq. \ref{eq-diff}, $\mathcal{F}^m(\rho_0)$ serves as an approximation of the target state and the other terms are differences between $ \mathcal{F}^m(\rho_0)$ and $\rho_x$. 

Let us first focus on the coefficient $C_m^x (\Gamma\delta t)^{m-x}$. Denote $x_{\operatorname{top}}$ satisfying $(m-x_{\operatorname{top}})/(x_{\operatorname{top}}+1)\Gamma\delta t=1$. Then this coefficient increases with the increasing of $x\in [0,x_{\operatorname{top}}]$ and decreases with the increasing of $x\in [x_{\operatorname{top}},m-1]$. Then we have the following remarks. 

\begin{enumerate}

\item For large $x$, $ \mathcal{F}^m(\rho_0) -\rho_x$ will become relatively small as we have shown in the main text for the reason that both the adjoint channel and the actual dissipative process will converge to the same steady state. However, the coefficients $C_m^x (\Gamma\delta t)^{m-x}$ may become very large, and the errors from finite sampling times and circuit approximations will be significantly amplified, which will affect the performance of this reconstruction procedure.  

\item For small $x$, $ \mathcal{F}^m(\rho_0) -\rho_x$ will not contribute to the total summation as the coefficients $C_m^x  (\Gamma\delta t)^{m-x} $ decreases exponentially with the decreasing of $x$. In this case, some finite truncation is sufficient to give accurate results. 

\end{enumerate}

Then, there is a trade-off between stability and accuracy during the reconstruction of the dissipated density matrix using Eq. \ref{eq-diff}. In our simulation, a sufficient large truncation is used to obtain reliable results. Furthermore, for the properties of the adjoint density matrix and dissipated density matrix in Eq. \ref{eq-steady} in the main text, we use the following reconstruction in our simulations. For the short-time simulation, we use the reconstruction procedure ($\mathcal{R}$) to recover the actual dynamical process. For long-time simulation, we use the adjoint channel to approximate the actual process directly. For an immediate time, we can suppress $\delta t$ to reduce the total error. In this case, when $\delta t$ is sufficiently small,  the reconstructed density matrix can serve as a good approximation of the truly time-evolved state.
%these two methods yield the same results. 

\section{Extra simulation results in the dissipative quantum $XY$ model}
\label{supp-plott10}

\begin{figure}[ht]
    \centering
    \includegraphics[width=0.6\linewidth]{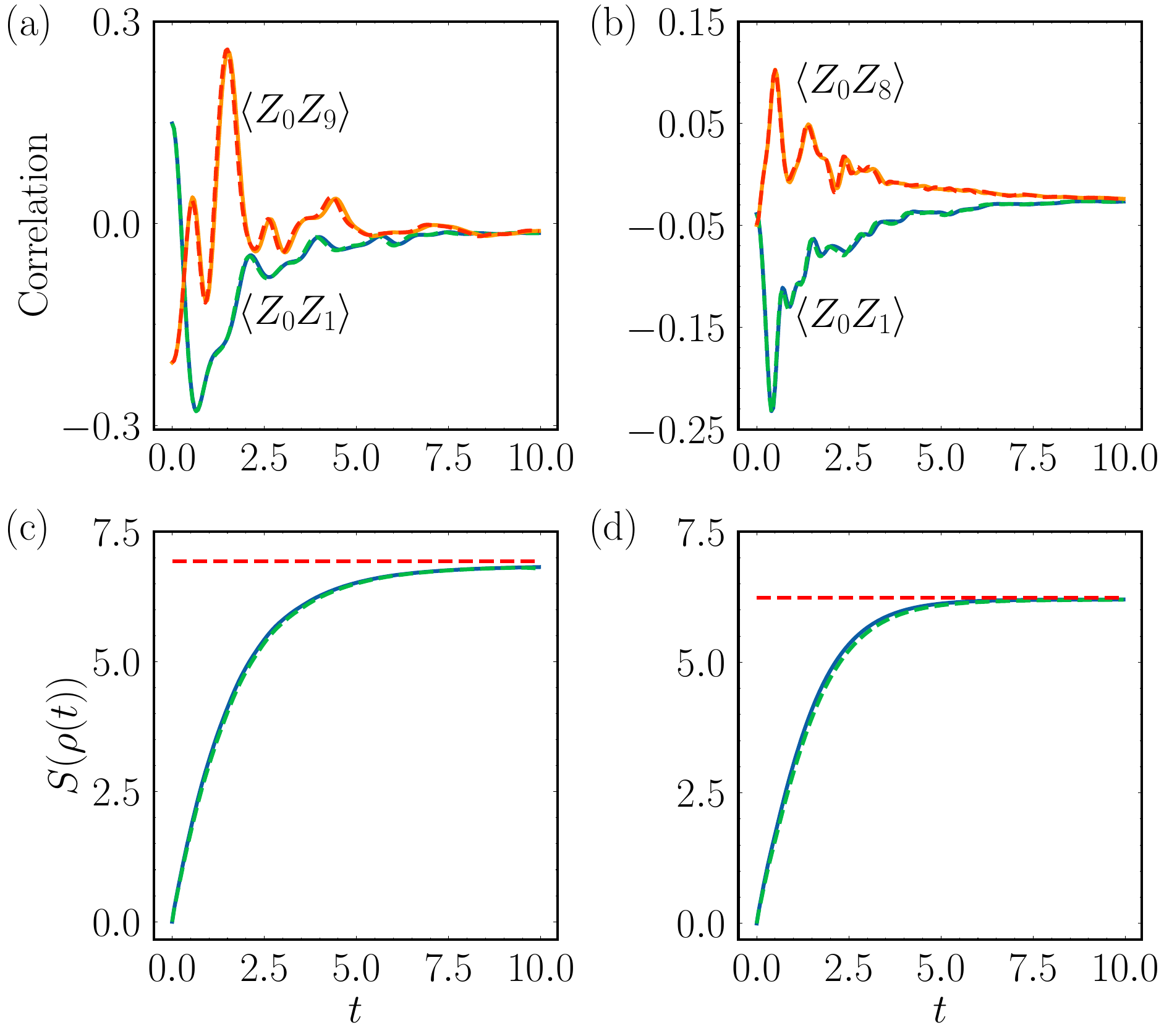}
    \caption{
    {\bf Simulation results for the dissipative quantum $XY$ model for $T=10$}. All the parameters are the same as that used in Sec. \ref{sec-xymodel}; and the meaning of the curves are the same as that in Fig. \ref{fig-xymodel}. At $t\to 10$, the thermalization is reached both from the correlation and entropy.}
    \label{figsupp_xy_10}
\end{figure}

In Sec. \ref{sec-xymodel} in the main text, we simulated the dissipative $XY$ model with simulation time $T=5$. To show that the system quickly reaches equilibrium, here we provide simulation results for a longer time at $T=10$. We still take it as a short-time dynamics, in which the reconstruction function in Eq. \ref{rr} is applied. All the other parameters during the simulation are the same as that in the main text. The results are presented in Fig. \ref{figsupp_xy_10}, showing the reach of thermalization at about $t\to 10$. Moreover, results obtained with our method agree with the exact dynamics, showing the reliability of our procedure. We have verified that in the long-time dynamics ($= 10$), the results from the dissipated density matrix and adjoint density matrix yield the same results. 

In Fig. \ref{figsupp_xy_10}, we have used time-step $\delta t = 0.05$, which is sufficient to obtain numerically reliable results. To verify the accuracy as a function of time-step, we next simulate the dissipative quantum $XY$ model with $\delta t$ varied from 0.01 to 0.1. We consider 
\begin{equation}
    |\Delta \langle Z_0Z_j\rangle | = |   \langle Z_0Z_j\rangle_{\operatorname{reconstruct}} -\langle Z_0Z_j\rangle_{\operatorname{exact}} |,
\end{equation}
where $j=1$ and $j=n-1$ represent short- and long-range correlations, respectively. The correlations with subscripts ``reconstruct" and ``exact" mean results were obtained using our method, and the results corresponded to the exact dynamics.

\begin{figure}[ht]
    \centering
    \includegraphics[width=0.6\linewidth]{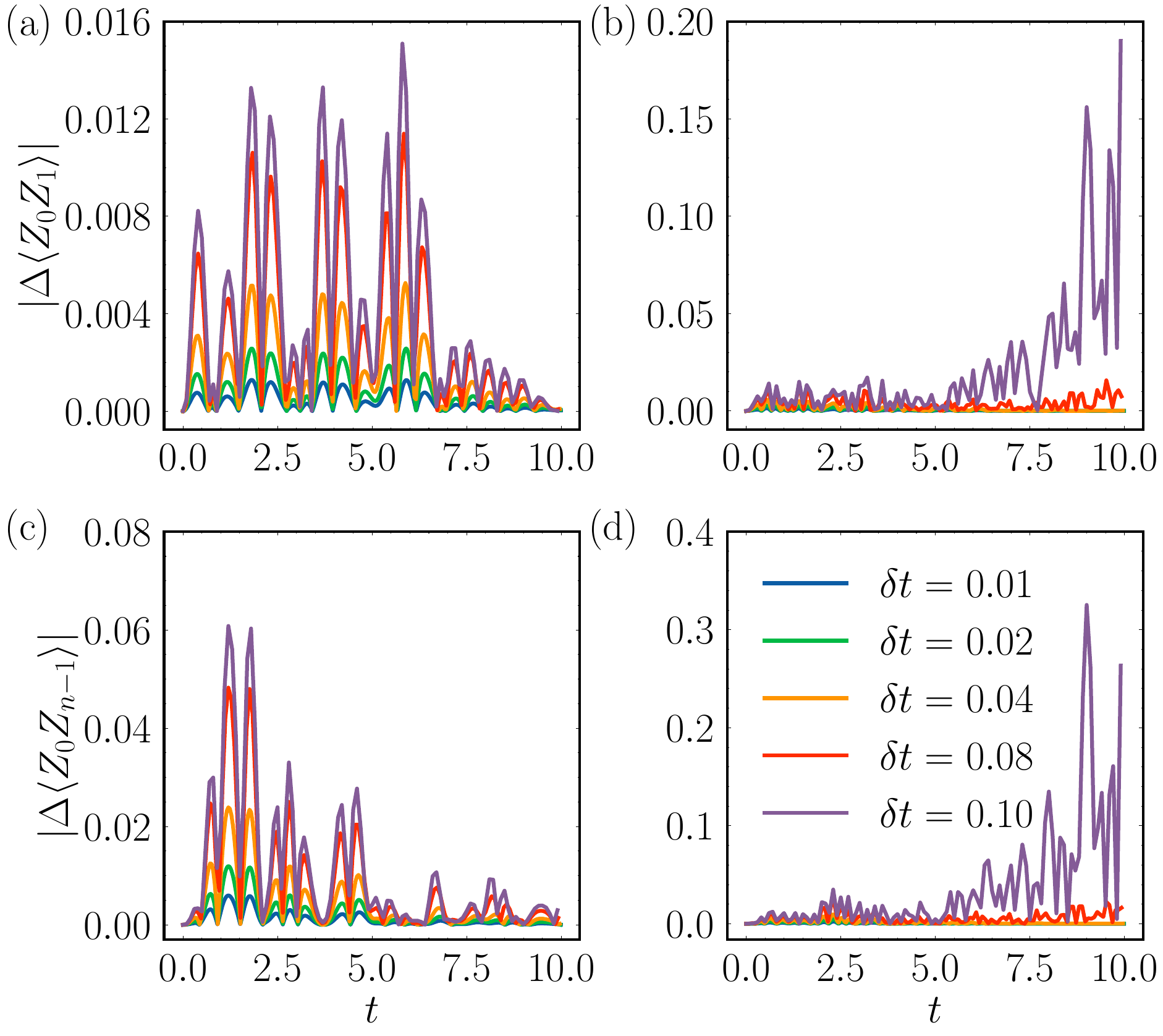}
    \caption{
    {\bf Simulation error with respect to different $\delta t$ in simulating the dissipative quantum $XY$ model.}
    The left and right two plots are the results for the one-dimensional model and two-dimensional model, respectively.
    }
    \label{figsupp_xy_1error_0}
\end{figure}

The result is shown in Fig. \ref{figsupp_xy_1error_0}. We can conclude: Firstly, simulations with $\delta t\leq 0.05$ work well in all models. Secondly, the error does not strictly increase with $T$ and is always smaller than $T\delta t$. In subfigures (b) and (d), the error increases quicker when $t>5.0$. This may caused by the instability in the reconstruction, and the data processing strategies discussed in Appendix \ref{supp-dataprocess} can be considered.

\section{Sampling of the adjoint channel}
\label{supp-samplemixedchannel}

Note that the number of Kraus operators of the adjoint channel $\mathcal{F}$ is $D+1$, expressed in a mixed-unitary form as
\begin{equation}
\mathcal{F}(\rho_0)= \sum_{\alpha=0}^D p_\alpha U_\alpha \rho_0 U_\alpha^\dagger.
\end{equation}
Based on this equation, we have
\begin{equation}
\begin{aligned}
    \mathcal{F}^2(\rho_0) =& \mathcal{F} \left(  \sum_{\alpha_1=0}^D p_{\alpha_1} U_{\alpha_1} \rho_0 U_{\alpha_1}^\dagger \right) \\
    =& \sum_{\alpha_2=0}^D p_{\alpha_2} U_{\alpha_2} \left(  \sum_{\alpha_1=0}^D p_{\alpha_1} U_{\alpha_1} \rho_0 U_{\alpha_1}^\dagger \right) U_{\alpha_2}^\dagger \\
    =& \sum_{\alpha_1=0}^D \sum_{\alpha_2=0}^D p_{\alpha_1}p_{\alpha_2} U_{\alpha_2}U_{\alpha_1} \rho_0 U_{\alpha_1}^\dagger U_{\alpha_2}^\dagger\\
    =& \sum_{\alpha_1=0}^D \sum_{\alpha_2=0}^D p_{\alpha_1}p_{\alpha_2} U_{\alpha_2}U_{\alpha_1} \rho_0 \left( U_{\alpha_2} U_{\alpha_1}\right)^\dagger.
\end{aligned}
\end{equation}
In a similar way to multi-step simulation, the composite quantum channel will become
\begin{equation}\label{mrho}
\mathcal{F}^m(\rho_0)  = \sum_{\bm{\alpha}} p_{\bm{\alpha}} U_{\bm{\alpha}} \rho_0 U_{\bm{\alpha}}^\dagger,
\end{equation}
where $\bm{\alpha} = \{  \alpha_1,\alpha_2,\cdots,\alpha_m \}$ is the index parameter with $\alpha_i\in\{ 0, 1,\cdots,D\},\forall i$.$p_{\bm{\alpha}} = p_{\alpha_1}p_{\alpha_2}\cdots p_{\alpha_m}$ and $U_{\bm{\alpha}} = U_{\alpha_m} \cdots U_{\alpha_2}U_{\alpha_1} $ are the corresponding probability and unitary operations. This process is resemblance to the evolution of state in path-integral formulism, in which the state $\mathcal{F}^m(\rho_0)$ is the summation of all possible paths. 

It can be seen that Eq. \ref{mrho} is a sum of $(D+1)^m$ terms, which grows exponentially with the number of evolution steps. We perform full computation to show the validity of our method in the numerical simulations. In real simulations with quantum chips, one can use strategies like the quantum trajectory method to approximate the effect of the quantum channel. 
% Moreover, the sign problem in Monto-Carlo samplings can be potentially avoided as probability $p_{\bm{\alpha}}$ are all positive. 
Therefore, A sufficient number of sampling times $M$ can provide accurate approximations based on the law of large numbers, where the error decreases as $1/\sqrt{M}$ according to the central limit theorem. This indicates that the $M$ does not need to increase exponentially with the number of steps, especially in the regular cases that $D\in \text{poly}(n)$, which have been studied in \cite{Dalibard1992Wave, carmichael2009open}.

To show that the quantum channel can be efficiently sampled, we provide additional tests here. We consider the adjoint channel derived from the 8-qubit one-dimensional dissipative quantum $XY$ mode, in which all model parameters are kept the same. According to Eq. \ref{eq-adjointchannel}. it is:
\begin{equation}
    \mathcal{F}(\rho) = \sum_{i=0}^n p_i U_i \rho U_i^\dagger,
\end{equation}
where:
\begin{equation}
    p_0 = \frac{1}{1+\Gamma\delta t},\quad
    U_0 = e^{-iH\delta t}, \quad
    p_k = \frac{\gamma\delta t}{1+\Gamma\delta t}, \quad
    U_k = Z_{k-1}, \quad
    k=1,2,\cdots,n.
\end{equation}
Starting from a random quantum state $\rho$, we consider the case of applying the quantum channel $20$ times and measure with the observable $O=Z_0Z_1$. The target value is $\operatorname{Tr}[ \mathcal{F}^{20} (\rho) O  ]$.

\begin{figure}[ht]
    \centering
    \includegraphics[width=0.6\linewidth]{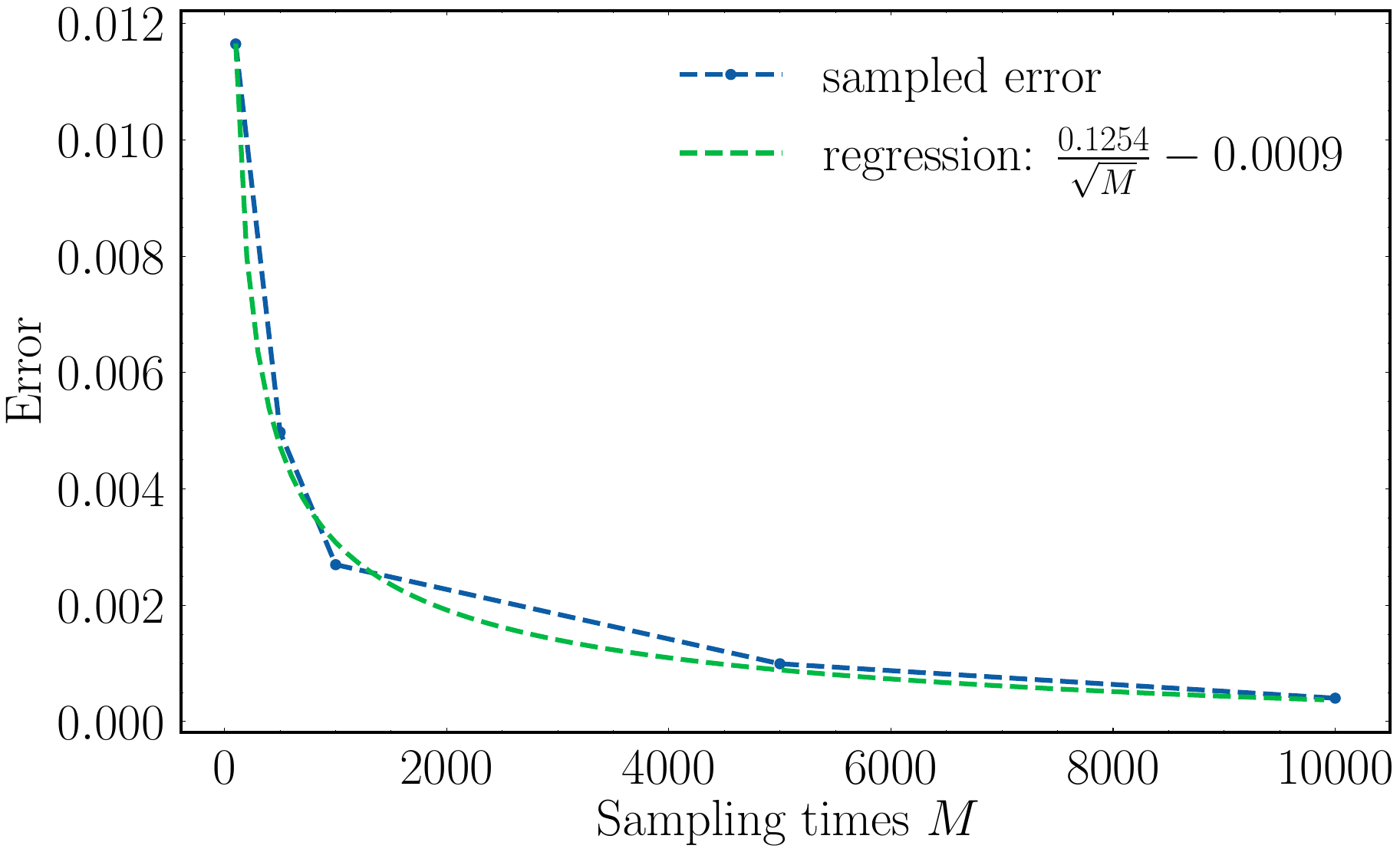}
    \caption{Sampling errors with respect to different sampling times. A function $e=a/\sqrt{M}+b$ is used to fit the data. The error decreases approximately at a rate of $1/\sqrt{M}$, which is in agreement with the central limit theorem.}
    \label{fig:sample}
\end{figure}

In this case, the total number of Kraus operators in the resulting channel is $9^{20}\approx 1.22\times 10^{19}$. We test the sampling error (absolute value) with respect to different sampling times $M$ (100, 500, 1000, 5000, 10000) and perform a fit using the function $e=a/\sqrt{M}+b$. The result is shown in Fig. \ref{fig:sample}.  From the results, we can conclude that the sampling error decreases approximately at a rate of $1/\sqrt{M}$, as dictated by the central limit theorem. We numerically find that about 1000 samples are sufficient to obtain a  result with error below $0.004$. This provides extra evidence that the channel can be efficiently sampled.

Finally, note that our method for simulating open quantum systems dynamics includes constructing the adjoint channel, sampling it, averaging all samples, and reconstructing the actual dynamics. This section analyzes and tests the effectiveness of sampling the adjoint channel. Other parts, such as averaging the $M$ samples and reconstructing the actual dynamics, can be efficiently executed on classical computers. Then the scalability of the simulation method is guaranteed.

\bibliography{ref.bib}
\bibliographystyle{quantum}
\end{document}